\documentclass{elsart}
\usepackage{epsfig}

\usepackage{graphicx}
\usepackage{amssymb}
\usepackage{amsmath}
\usepackage{times,mathptm}
\newcommand{\imai}{{\rm i}}  
\begin{document}

\begin{frontmatter}
Proceedings of HCIS-12 (2001), to appear in Physica B

\title{Gain and Loss in Quantum Cascade Lasers}
\author{A. Wacker and S.-C. Lee}

\address{Institut f\"ur Theoretische Physik, Technische Universit\"at Berlin, 10623 Berlin, Germany}

\begin{abstract}
We report gain calculations for a quantum cascade laser
using a fully self-consistent quantum mechanical
approach based on the theory of nonequilibrium Green functions.
Both the absolute value of the gain as well as the spectral
position at threshold are in excellent agreement with 
experimental findings for $T=77$K.
The gain strongly decreases with temperature.	
\end{abstract}

\begin{keyword}
quantum cascade lasers \sep
gain \sep
quantum transport

\PACS 42.55.Px \sep 05.60.Gg \sep 73.40.-c
\end{keyword}

\end{frontmatter}

\section{Introduction}

Quantum cascade lasers (QCL) \cite{qcl} 
constitute an important source of
midinfrared radiation. These complex
semiconductor heterostructures exhibit
a sequence  of several ($\sim 30$) periods. 
Each period consists of an active region, where the 
optical transition takes place, and
an injector region guiding the electrons from the lower level
into the upper level of the active region in the neighboring period
in a certain bias range.

The operation of QCLs
is based on a complex interplay between tunneling through injector regions,
scattering transitions, and the interaction with the
infrared light. While semiclassical theories based on the
Boltzmann equation \cite{theory} can explain several features 
of the relaxation processes in the active region, 
a quantum transport treatment is indispensable to study
the functionality of the injector as well as its interplay with the
active region. Such transport simulations have been
performed very recently applying nonequilibrium Green functions \cite{WAC01a,LEE01}
and density matrix theory \cite{IOT01a} for the QCL.
Besides the transport properties, the gain spectrum is also strongly affected
by scattering processes. While these effects are commonly taken into
account by means of  phenomenological broadening,
we present a full
self-consistent quantum mechanical description here.

We model the transport in QCLs employing the technique
of nonequilibrium Green functions \cite{HAU96}.
The quantum  kinetic equations are solved using self-energies in the
self-consistent Born approximation for phonon and impurity scattering.
The interaction with the microwave field is treated within
linear response with respect to the {\em nonequilibrium state}.
In this way, we obtain the complex dielectric susceptibility
and the gain coefficient $g$. This provides us with a self-consistent
formalism combining scattering and coherent evolution,
which provides level occupations and broadening of optical spectra
within a single approach.
As an example we present results for the
GaAs/AlGaAs structure of Ref.~\cite{SIR98}. 

\section{The model}
We set up our Hamiltonian within the basis of Wannier-functions, 
which allows for a direct calculation from nominal sample parameters using
standard envelope functions (see \cite{WAC01} for details).
The self-energies for phonon and interface roughness scattering 
are taken into account within the self-consistent Born approximation.
The solution of the Kadanoff-Baym equations \cite{HAU96}
provides us with the lesser and retarded Green functions
\begin{equation}
{\bf G}^{</{\rm ret}}({\bf k},t_1,t_2)=
\int\frac{\d E}{2\pi}
{\bf G}^{</{\rm ret}}({\bf k},E)e^{-\imai E(t_1-t_2)/\hbar}
\label{EqGneq}
\end{equation}
in the nonequilibrium state (a stationary state, thus the Green functions
only depend on the time difference $t_1-t_2$).
Here bold-face capital letter describe matrices
$G^{\mu\nu}_{mn}$, where $\nu$ is the band index and $n$ the period index 
in the QCL of the respective
Wannier functions $\Psi_n^{\nu}(z)$. ${\bf k}$ denotes a two-dimensional
wave vector, describing the free-particle behavior in the plane
perpendicular to the growth direction.
Details are given in Ref.~\cite{LEE01}, where this approach was applied
to the calculation of the current-field relation in
good agreement with the experimental values.

Now we consider the coupling to an external optical field $F$ 
(in $z$ direction)
with the vector potential
$\vec{A}(t)=\frac{1}{\imai\omega}\int \frac{\d \omega}{2\pi}
F(\omega)e^{-\imai\omega t} \vec{e}_z$
which provides us with the perturbation potential
$\delta\hat{U}(t)=-\frac{e}{m_e}\vec{A}\cdot \hat{\vec{p}}$
in the Hamiltonian (neglecting quadratic terms).
Its matrix representation reads
$\delta {\bf U}(t)=\int \d \omega/(2\pi)
\delta {\bf U}(\omega)e^{-\imai\omega t}$
with
\begin{equation}
\delta U^{\nu\mu}_{n,m}(\omega)\;=\;
-\frac{e}{\hbar\omega} F(\omega)
\langle \Psi^{\nu}_n|\frac{\hbar}{\imai}\frac{\hat{p}_z}{m_e}
|\Psi^{\mu}_m\rangle.
\end{equation}
Here $\hat{p}$ is
the momentum operator and $m_e$ is the electron mass.

We linearize the time-dependent Kadanoff-Baym equations
with respect to
$\delta {\bf U}(t)$ around the nonequilibrium state (\ref{EqGneq}).
Applying the Fourier representation
\begin{equation}
\delta {\bf G}({\bf k},t_1,t_2)\;=\;
\int \frac{\d \omega}{2\pi} e^{-\imai\omega t_1}
\int\frac{\d E}{2\pi}
\delta{\bf G}({\bf k},\omega,E)e^{-\imai E(t_1-t_2)/\hbar},
\end{equation}
the field-induced change of Green functions is given by
\begin{equation}\begin{split}
\delta {\bf G}^{<}({\bf k},\omega,E)\;=&\;
{\bf G}^{\rm ret}({\bf k},E+\hbar \omega)\delta {\bf U}(\omega)
{\bf G}^{<}({\bf k},E)\\
\;+&\;{\bf G}^{<}({\bf k},E+\hbar \omega)\delta {\bf U}(\omega)
{\bf G}^{\rm adv}({\bf k},E).
\label{EqGlessLinResp}
\end{split}
\end{equation}
In general, the self-energies also change
due to the perturbation, which gives more terms of the same order.
This corresponds to ladder corrections, which are 
neglected here.

From the lesser Green function we obtain the change in current
density:
\begin{equation}
\begin{split}
\delta J(t)\;=&\;\frac{e}{LA}\delta\langle 
\frac{\hat{p}_z}{m_e}\rangle\\
=&\;\frac{2\mbox{(for spin)}\,e}{dA\imai} \sum_{{\bf k},n,\mu,\nu}
\langle \Psi^{\nu}_0|\frac{\hat{p}_z}{m_e}|\Psi^{\mu}_n
\rangle
\delta G^{\mu\nu <}_{n,0}({\bf k};t,t)
\end{split}
\end{equation}
where scattering induced currents have been neglected.
Here $A$ is the cross section, $L$ is the total length, and
$d$ is the period of the sample.
The Fourier transform $\delta J(\omega)$ provides us with
the Polarization $\delta P(\omega)=\imai \frac{\delta J(\omega)}{\omega}$
and the complex susceptibility 
$\chi (\omega)=\frac{P(\omega)}{\epsilon_0 F(\omega)}$.
From standard electrodynamics (see, e.g., Sec. 7.5 of \cite{JAC98a})
we obtain the gain coefficient (for low absorption or gain)
\begin{equation}\begin{split}
g(\omega)\approx - \frac{\omega}{c}\frac{\Im\{\chi(\omega)\}}
{\sqrt{\epsilon_r+\Re\{\chi(\omega)\}}}
\end{split}
\end{equation}
where $\epsilon_r=13$ is the static dielectric susceptibility of GaAs.
Note that all quantities entering the calculation are material parameters.

\section{Results}

\begin{figure}
\noindent
\epsfig{file=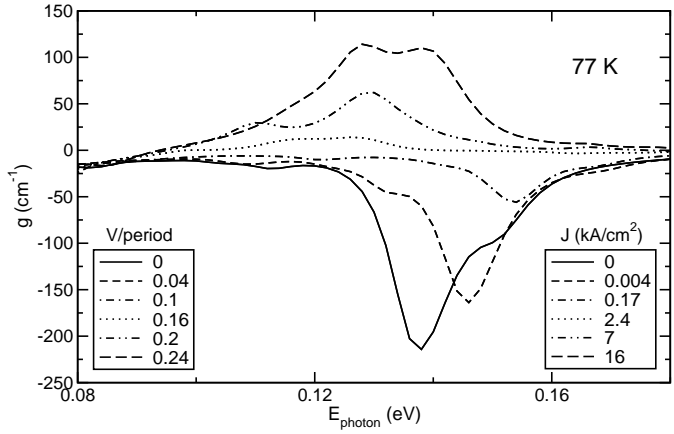,angle=-90,width=10cm}\\[0.2cm]
\caption{Calculated gain and absorption ($g<0$) spectra
for the QCL from \cite{SIR98}.
Data for different voltage drops per period and corresponding
current densities are displayed.}\label{Figgain77}
\end{figure}

\begin{figure}
\epsfig{file=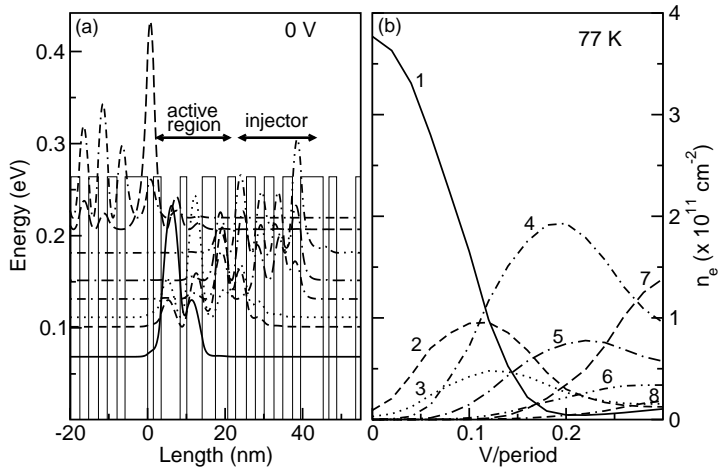,angle=-90,width=10cm}\\[0.2cm]
\caption{(a) Square of Wannier functions associated with one period
together with the band alignment. 
(b) Population of the Wannier levels shown in (a)
as a function
of voltage drop for $T=77$ K. The curves are labelled according to
increasing energy, i.e. curve 1 shows the population of the
Wannier level with the lowest energy.}
\label{FigWan0pop77}
\end{figure}

We have applied the above formalism to 
the GaAs/AlGaAs structure reported in Ref.~\cite{SIR98}.
Fig.~\ref{Figgain77} shows absorption and gain spectra at 77 K calculated
for different applied bias, ranging from 0 to 0.24 V/period
(0.1 V/period $\sim$ 22 kV/cm). At zero bias, there is only
absorption with a strong absorption line at around 138 meV.
As the applied voltage increases there is an overall decrease
in absorption and a blue
shift of the absorption peak. At around 0.14 V/period or 1 kA/cm$^2$,
there is a transition to gain at around 120 meV. From then on, 
as the voltage increases, the gain increases further, accompanied
by a shift of the peak gain to higher frequencies. A double peaked
structure also appears in the gain structure. At present, we have
not yet made a detailed analysis of the spectra, and we cannot
definitely assign the absorption or gain features, and shifts 
in these features, to specific transitions in the structure. 
We can, however, obtain some understanding of the transition from 
absorption to gain if we consider the populations of the Wannier
levels. Fig.~\ref{FigWan0pop77}(a) shows the Wannier levels in one period of the structure,
and Fig.~\ref{FigWan0pop77}(b) 
shows the populations in each of these levels as the
voltage is increased. At zero bias, all the population is in the
lowest Wannier level which is localized in the active region. 
As the voltage increases, the population moves out of this level 
into the next two higher levels which are partially in the active
region and partially in the injector, and with yet further voltage
increase, the population moves into the fourth and fifth levels
which are localized mainly in the injector. Finally, above 0.25 V/period
we see a large population inversion with the population lying mainly
in injector levels and in the highest Wannier level in the active region,
and with almost no population in the lowest level. The transition
from absorption to gain occurs between 0.15 to 0.2 V/period. This is also
the point where the population in the excited level in the active region
first exceeds the population in the lowest level. Thus, with this
picture we can follow the movement of the carriers across the period
from the lowest level in the active region, through the injector,
and into the highest level in the active region as the applied voltage
increases. This movement can be related to the shift and realignment
of the Wannier levels as the voltage changes.

\begin{figure}  
\epsfig{file=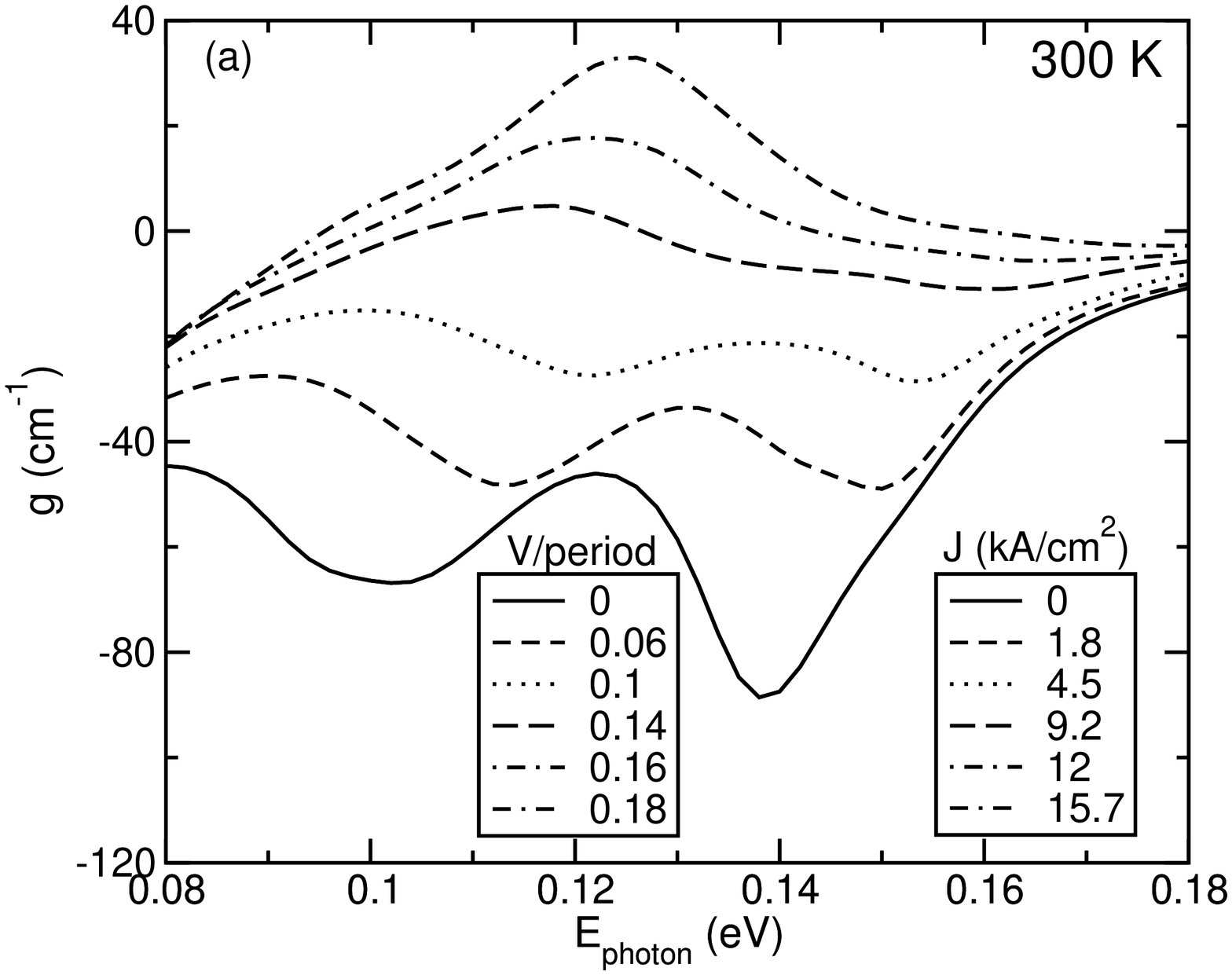,angle=-90,width=7cm}
\epsfig{file=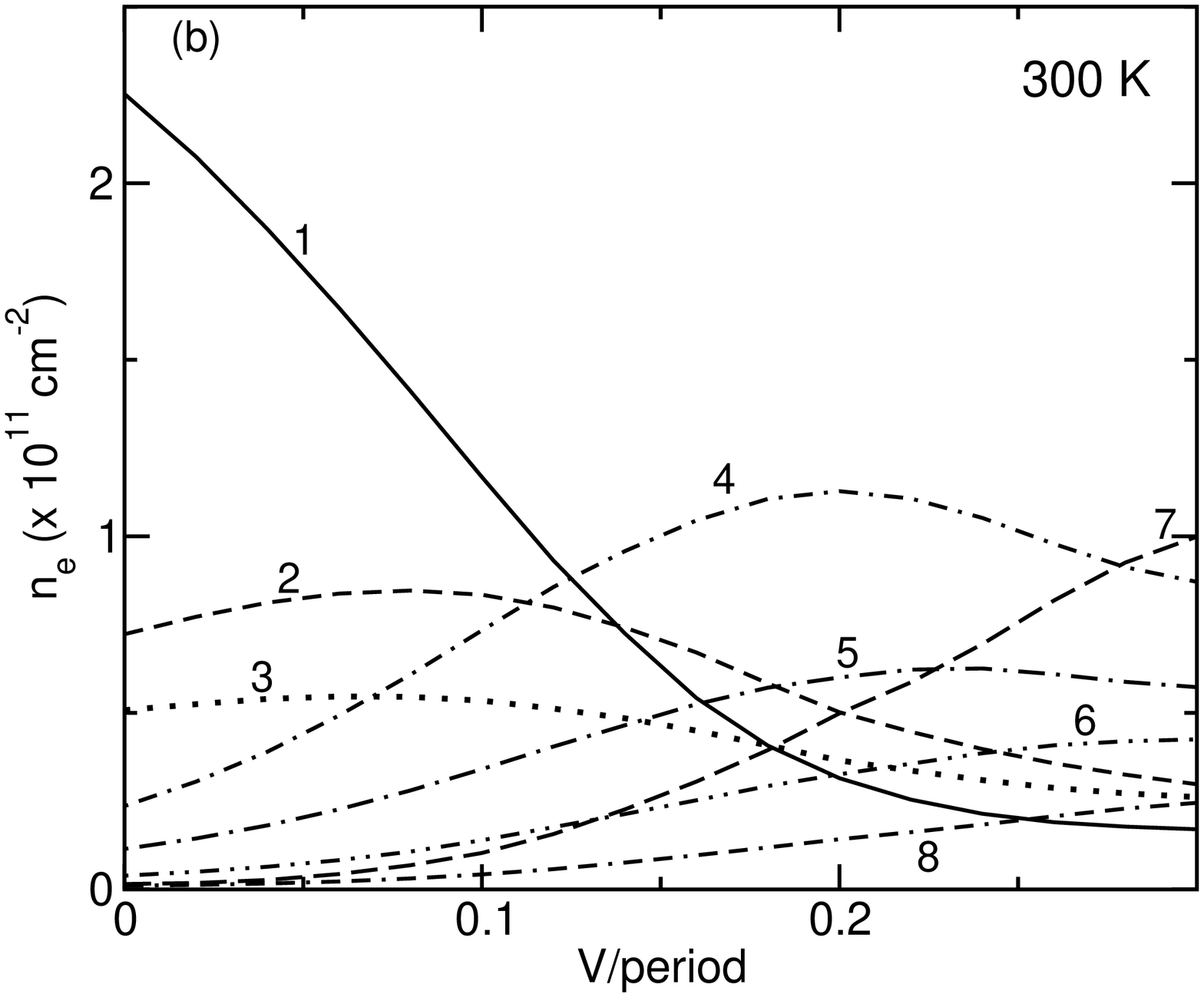,angle=-90,width=7cm}\\[0.2cm]
\caption{(a) Calculated gain and absorption ($g<0$) spectra
for 300 K; (b) population of the Wannier levels as a function
of voltage drop. The curves are labelled as  in
Fig. 2(b). }
\label{Figgain300}
\end{figure}

In Fig.~\ref{Figgain300}(a), we show gain and 
absorption spectra calculated for 300 K.
The main differences compared with the spectra at 77 K, are that
the absorption and gain are reduced  by a factor of 
more than two at 300 K, compared with their values at 77 K, for the
same current density. The width of the spectra is also wider at
the higher temperature. Both these effects can be understood if
we consider the populations of the Wannier levels at 300 K shown in
Fig.~\ref{Figgain300}(b). The figure shows that
the population is much more widely distributed over the different levels 
at higher temperature.
There is a significant proportion of the population in higher levels
even for zero bias, unlike the case at 77 K where the population
is mainly in the lowest level. Similarly, in the case of population
inversion at higher voltage, there is a larger fraction of the
population in the lowest level at 300 K than at 77 K. Hence, because
the population difference between different levels is reduced at
higher temperature, this causes a reduction in the absorption or gain.
The redistribution of carriers over more levels allows more
transitions to occur and this, together with the increased
phonon scattering, explains the larger width of the
spectra at 300 K compared to 77~K.

\section{Discussion}
We have presented gain calculations for the QCL of Ref.~\cite{SIR98}
using a fully self-consistent quantum mechanical
approach on the basis of nonequilibrium Green functions.
We find that gain sets in for current densities around 1 kA/cm$^2$
and increases with current. For $T=$ 77~K, the peak gain coefficient is
60 cm$^{-1}$ at 7 kA/cm$^2$ and a photon
energy of 130 meV, in excellent agreement
with the findings of Ref.~\cite{SIR98}
(material gain $G_M=63$~cm$^{-1}$ at threshold $J_{\rm th}$=7.2 kA/cm$^2$).
The spectral position of the gain spectrum agrees
well with the findings of Ref.~\cite{EIC00} although neither
a double peak structure nor a blue shift was resolved
in the experiment.
For room temperature, the gain is significantly reduced
and becomes less than the estimated loss 
explaining the lack of lasing operation.

For practical reasons, 
the influence of electron-electron scattering has been
neglected in these calculations. While the current density
seems not to depend strongly on the type of scattering
mechanism (see the discussion in \cite{LEE01}), 
electron-electron interaction may affect the 
gain spectrum. This constitutes an important issue
for future research.

We thank  A. Knorr and M. Pereira for helpful discussions.
Financial support by DFG through FOR394 is gratefully
acknowledged.

\end{document}